# Passivity of Metals: Potential-Step Transients for Passive Current Density and Barrier Layer Thickness in the Point Defect Model

Bosco Emmanuel, CSIR-CECRI, Karaikudi-630006, India

**Abstract**

In an earlier pre-print [1] we developed a variant of the point defect model that corrected a flaw in one of the defect reactions and to be specific reaction 3 of the original point defect model [2] was replaced by reaction 3' of the variant. Here we apply the corrected model to find the time evolution of the passive current density and the barrier layer thickness. Though the functional forms agree with the forms reported earlier by Macdonald and co-workers [3], the composition of the parameters differ significantly. This explains the experimental success of the flawed model. This semblance is also manifested in the diagnostics for the current transient involving the time-derivative of the current density. The present theory will be useful for the correct analysis of transients which result from switching the potential in the anodic or cathodic direction from an initial steady state.

**Introduction**

The general time-dependent expression for the current density $i$ in the point defect model is:

$$i_{ss} = F.\{-\delta.J_M + \delta.J_i + 2.J_O + (\delta - \chi).R_7\} \tag{1}$$

We make two assumptions: (i) the evolution of the transient state is not too fast and hence even in a transient situation the rate of production of the metal vacancy $V_m$ by reactions 1 and 2 equals the rate of its annihilation by reaction 3' and (ii) the barrier layer is so thin that the transport of defects in the barrier layer is in a steady state; otherwise one needs to solve the full moving boundary value problem (the Stephen's problem). These two assumptions imply

$$J_i - J_M = \left(\frac{q}{(\chi/2)}\right).J_O \tag{2}$$

$$J_i = k_2 \tag{3}$$

$$J_M = -k_4 \tag{4}$$

Using these equations in (1) we obtain

$$i = F.\{(\delta + \frac{\chi}{q}).(k_2 + k_4) + (\delta - \chi).R_7\} \tag{5}$$

Now

$$\frac{dL}{dt} = \Omega.\{R_3 - R_7\} \tag{6}$$

Where $R_3$ and $R_7$ are rate of formation and destruction of the oxide $MO_{\chi/2}$ at $x=0$ and $x=L$ respectively.

$$R_3 = \frac{J_O}{(\chi/2)} \tag{7}$$

Use equations (2) to (4) in equation (7) to obtain

$$\frac{dL}{dt} = \Omega \cdot \{\frac{k_2 + k_4}{q} - R_7\} \tag{8}$$

Differentiating equation (5) with respect to time

$$\frac{di}{dt} = F \cdot \{(\delta + \frac{\chi}{q}) \cdot (\frac{dk_2}{dt} + \frac{dk_4}{dt}) + (\delta - \chi) \cdot \frac{dR_7}{dt}\} \tag{9}$$

At $t=0$ the system is assumed to be in a steady state at an applied initial potential $V_i$ from which the potential is switched to the final potential $V_f$. During this switching the quasi-steady state approximation made above will not hold and besides there will be a capacitance spike. However for $t$ sufficiently greater than the rise time of the potentiostat and the time required to establish the quasi-steady state, and if this time is denoted $0+$, equation (9) may safely be integrated from $0+$ to $t$ yielding

$$i(t) - i(0+) = F \cdot (\delta + \frac{\chi}{q}) \cdot k_2^0 \cdot \exp(a_2 V_f + c_2 pH) \cdot \{\exp(-b_2 L(t)) - \exp(-b_2 L(0+))\} \tag{10}$$

Note that we have used the forms for $k_2, k_4$ and $R_7$ as originally proposed by Macdonald and co-workers [2] as these are amenable to analytic solutions. The terms involving $k_4$ and $R_7$ have naturally dropped out of equation (9).

Here it is reasonable to assume that $L(0+) = L_{ss}^0$ which is the initial steady state barrier layer thickness before switching the potential while we can NOT set $i(0+) = i_{ss}^0$ for reasons mentioned below equation (9). However there is a way to find $i(0+)$ as shown in the end of this paper.

**Analytic Expression for the Barrier Layer Thickness $L(t)$**

Rewrite equation (8) as

$$\frac{dL}{dt} = \frac{\Omega}{q} k_2 + (\frac{\Omega}{q} k_4 - \Omega \cdot R_7) \tag{11}$$

$$= A \exp(-b_2 L(t)) + C \tag{12}$$

Where

$$A = \frac{\Omega}{q} k_2^0 \exp(a_2 V_f + c_2 pH) \tag{13}$$

And

$$C = \frac{\Omega}{q} k_4 - \Omega R_7 \tag{14}$$

where $C$ can be shown to be negative. $b_2$ is as defined by Macdonald et al.

Integrating equation (12) we obtain

$$L(t) = L_{ss}^0 + (1/b_2) \ln\{1 + (A/C)\exp(-b_2 L_{ss}^0).(1 - \exp(-b_2 Ct))\} + Ct \tag{15}$$

It is interesting to note the functional similarity of $L(t)$ with the corresponding result reported by Macdonald et al [3]. Nonetheless it is important to point out that the composition of the parameters $A$, $C$ and $b_2$ is different from that of their counterparts a', c and b.

**Analytic Expression for the Current Density $i(t)$**

Using equation (15) in equation (10) we obtain

$$i(t) = i(0+) + F(\delta + \frac{\chi}{q}) k_2^0 \exp(a_2 V_f + c_2 pH).\exp(-b_2 L_{ss}^0). \left( \frac{1}{\left\{1 + \frac{A}{C}\exp(-b_2 L_{ss}^0)\right\}\exp(b_2 Ct) - \frac{A}{C}\exp(b_2 L_{ss}^0)} - 1 \right) \tag{16}$$

Here again the functional form is the same as that reported by Macdonald et al [3] though the composition of the parameters are different. Hence it is not surprising that the original point defect model even with its flawed reaction 3 has been successful in explaining experimental data.

**Diagnostics for the Current Transient**

Macdonald et al provided a diagnostic for the current transient in terms of the time derivative of the current density. In this section we derive a similar diagnostic for the corrected point defect model.

From equation (5)

$$\frac{k_2 + k_4}{q} = \frac{(i/F) - (\delta - \chi).R_7}{\chi + \delta.q} \tag{17}$$

Use this result in equation (8) to obtain

$$\frac{dL}{dt} = \Omega.\frac{i - \delta.F.R_{PB}.R_7}{F.(\chi + \delta.q)} \tag{18}$$

For $t > 0$ equation (12) becomes

$$\frac{dL}{dt} = \Omega . \frac{i - i_{ss}^f}{F.(\chi + \delta.q)} \qquad (19)$$

where $i_{ss}^f$ is the final steady state current density after the potential switching.

Use equations (10) and (19) in equation (9) to obtain

$$-\frac{\frac{di}{dt}}{[i(t) - i_{ss}^f]} = \frac{b_2 \Omega}{F(\chi + \delta q)}[i(t) - i(0+)] + b_2(\Omega/q)k_2^0 \exp(a_2 V_f + c_2 pH)\exp(-b_2 L_{ss}^0) \qquad (20)$$

The similarity and the contrast are to be noted against the corresponding result from the original point defect model [2].

A plot of the left hand side of equation (20) versus $[i(t) - i(0+)]$ should yield a straight line.

**Relation between the initial and the final steady state current densities**

Lastly we promised to find $i(0+)$. Taking the steady state limit for the current density in equation (16), it is simple to show:

$$i(0+) = \frac{(\delta - \chi)}{\delta . R_{PB}} i_{ss}^f + F(\delta + \frac{\chi}{q}).[k_4 + k_2^0 \exp(a_2 V_f - b_2 L_{ss}^0 + c_2 pH)] \qquad (21)$$

Thus there is an interesting linear relationship between the initial current density and the final steady state current density.

**Conclusions**

**The difference between the predictions of the original point defect model and the present variant lie essentially in the composition of the parameters as far as the transients of the current density and the barrier layer thickness are considered. The theory and analysis of the EIS response of the two models should throw more light on these parametric differences to which we plan to turn next.**